\documentclass[epj,twocolumn]{webofc}
\usepackage[varg]{txfonts} % Web of Conferences font
\usepackage{amsmath,amssymb}
\usepackage{slashed}
\allowdisplaybreaks

\newcommand{\st}{{\scriptscriptstyle T}}

\DeclareMathOperator{\tr}{Tr}

\wocname{EPJ Web of Conferences}
\woctitle{TRANSVERSITY 2014}
\begin{document}
\title{Universality of TMD correlators}
\author{
M.G.A. Buffing\inst{1}\fnsep\thanks{\email{m.g.a.buffing@vu.nl}} \and
A. Mukherjee\inst{2}\fnsep\thanks{\email{asmita@phy.iitb.ac.in}} \and
P.J. Mulders\inst{1}\fnsep\thanks{\email{p.j.g.mulders@vu.nl}}
}
\institute{Nikhef and Department of Physics and Astronomy, VU University Amsterdam,\\
De Boelelaan 1081, NL-1081 HV Amsterdam, the Netherlands 
\and
Department of Physics, Indian Institute of Technology Bombay, Powai, Mumbai 400076, India
}
\abstract{In a high-energy scattering process with hadrons in the initial state, color is involved. Transverse momentum dependent distribution functions (TMDs) describe the quark and gluon distributions in these hadrons in momentum space with the inclusion of transverse directions. Apart from the (anti)-quarks and gluons that are involved in the hard scattering process, additional gluon emissions by the hadrons have to be taken into account as well, giving rise to Wilson lines or gauge links. The TMDs involved are sensitive to the process under consideration and hence potentially nonuniversal due to these Wilson line interactions with the hard process; different hard processes give rise to different Wilson line structures. We will show that in practice only a finite number of universal TMDs have to be considered, which come in different linear combinations depending on the hard process under consideration, ensuring a generalized universality. For quarks this gives rise to three Pretzelocity functions, whereas for gluons a richer structure of functions arises.}
\maketitle
\section{Introduction}
\label{s:intro}
In the description of hadronic scattering processes, one has to consider both hard scattering contributions as well as parton distribution functions (PDFs) that describe the hadrons initiating the interactions. We consider transverse momentum dependent PDFs (TMDs) by including transverse directions in momentum space in the description of these objects~\cite{Ralston:1979ys}. New phenomena appear and manifest themselves for example in the form of angular correlations between the particles involved in the process. Another effect is the sensitivity to polarization modes of the hadron and constituent partons that would not have been possible without the inclusion of these transverse directions. It is therefore relevant to study these TMDs.

In these proceedings, which are based on the Refs.~\cite{Buffing:2012sz,Buffing:2013kca}, we focus on the universality properties of these TMDs. In a color gauge invariant description, gauge links, path ordered exponentials, have to be included in the definition of TMDs. These gauge links appear as a result of gluon emissions coupling to the (colored) particles in the hard scattering process. It is this interplay between gauge links and the hard process that introduces a sensitivity and potential process dependence of the TMDs to the process in which it appears, since the gauge link structure is process dependent itself. We refer to Ref.~\cite{Bomhof:2006dp} for a tabulation of the structures. The Sivers effect is a consequence of the presence of different gauge link structures in different processes~\cite{Sivers:1989cc}. In turn, this warrants a study to make a classification of all the TMD structures that appear in the various processes, investigating the existence of a more generalized form of universality. In Section~\ref{s:quarks} we outline the generalized universality for quarks, published in Ref.~\cite{Buffing:2012sz} and in Section~\ref{s:gluons} we focus on the generalized universality for gluons, which has been published in Ref.~\cite{Buffing:2013kca}. In Section~\ref{s:Conclusions} we present some general conclusions and a brief discussion of the results.

\section{Quarks}
\label{s:quarks}
\begin{figure}[!tb]
\centering
\includegraphics[width=3.8cm,clip]{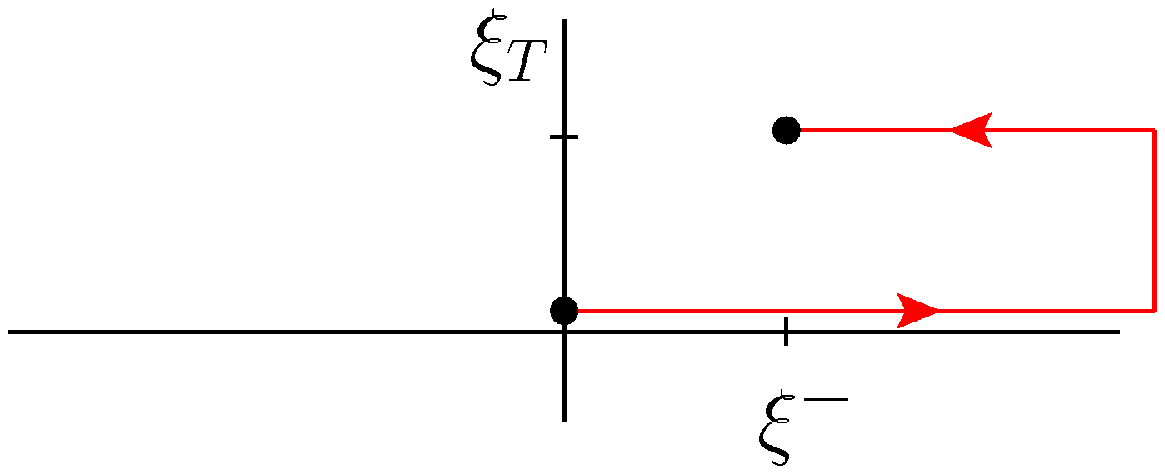}
\hspace{0.3cm}
\includegraphics[width=3.8cm,clip]{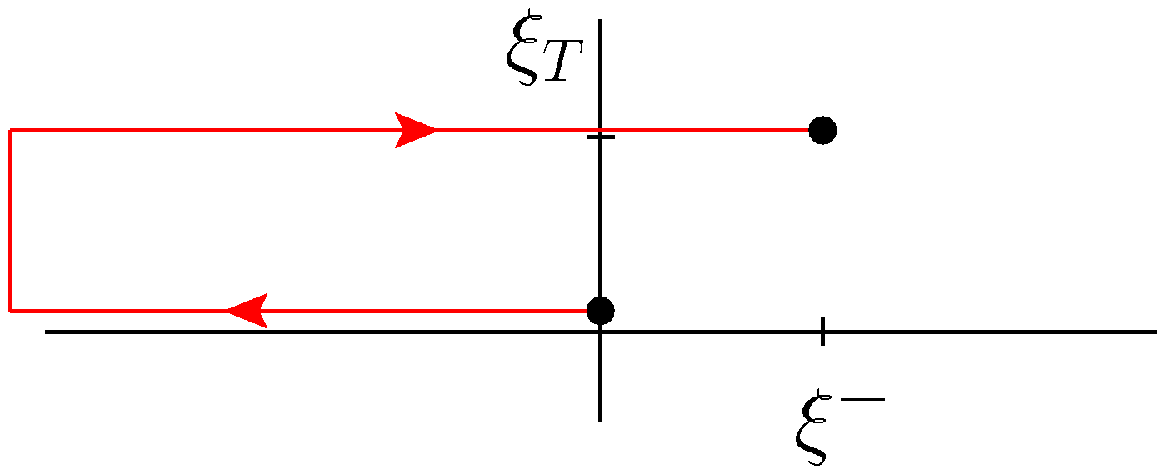}
\\[0.1cm]
\begin{small}
(a)\hspace{4.0cm} (b)
\end{small}
\\[0.1cm]
\caption{The two simplest gauge links for quark distribution functions. The dots indicate the positions $0$ and $\xi$ of the two quark fields in the correlator, while the path of the gauge link is indicated by the line connection the two positions. In the simplest configuration the gauge link runs through either plus or minus light cone infinity, illustrated in (a) and (b) respectively.
Figures taken from Ref.~\cite{Buffing:2012sz}.}
\label{f:GL_quarks}
\end{figure}

For quarks, the matrix element describing the correlator is given by
\begin{eqnarray}
\hspace{-5mm} \Phi_{ij}^{[U]}(x,p_{\st};n)&=&\int \frac{d\,\xi{\cdot}P\,d^{2}\xi_{\st}}{(2\pi)^{3}}\,e^{ip\cdot \xi} \nonumber \\
&&\hspace{0.1cm}\times\langle P\vert\overline{\psi}_{j}(0)\,U_{[0,\xi]}\psi_{i}(\xi)\vert P\rangle\,\Big|_{\xi\cdot n=0},
\end{eqnarray}
which contains a bilocal combination of quark fields connected by a gauge link $U_{[0,\xi]}$. This gauge link, ensuring color gauge invariance in the process, consists of a path ordered exponential. As will be explained later, the path depends on the process under consideration and is constructed out of staple like pieces, running through light cone infinity. They are of the form $U_{[0,\xi]}^{[\pm]} = U_{[0,\pm \infty]}^{[n]}U_{[0_\st,\xi_\st]}^{T}U_{[\pm \infty,0]}^{[n]}$, with $n$ being the direction along the light cone and $T$ the direction in the transverse plane. The two simplest paths are indicated in Fig.~\ref{f:GL_quarks}, connecting the fields through either plus or minus light cone infinity. These gauge links emerge due to soft gluon emission from the quark correlator coupling to the particle involved in the hard process. Initial state interactions (ISIs) give rise to minus gauge links and final state interactions (FSIs) imply plus gauge links.

The second way to describe the correlator is by writing an expansion in terms of transverse momentum dependent parton distribution functions (TMDs). The contributions for an unpolarized hadron are given by
\begin{equation}
\Phi^{[U]}(x,p_{\st};n)=\bigg\{f_1^{[U]}(x,p_\st^2)+ih_1^{\perp [U]}(x,p_\st^2)\,\frac{\slashed{p}_{\st}}{M}\bigg\}\frac{\slashed{P}}{2},
\label{e:quarkpar}
\end{equation}
with $h_1^{\perp [U]}(x,p_\st^2)$ being the Boer-Mulders function, the function describing transversely polarized quarks in an unpolarized proton, whereas $f_1 (x,p_\st^2)$ describes the unpolarized quark in an unpolarized proton. By including linearly or transversely polarized hadrons more TMDs have to be included in the parametrization, for which we refer to Ref.~\cite{Bacchetta:2006tn}.

As of now, we have two descriptions, which should be related to each other. In order to do so, we use transverse moments, weightings with transverse momenta, a procedure which can be applied at the level of both the TMDs and the matrix elements. For the matrix elements, the result of a single transverse weighting is given by~\cite{Boer:2003cm}
\begin{eqnarray}
\Phi_{\partial}^{\alpha[U]}(x) &\equiv & \int d^2 p_{\st}\,p_{\st}^{\alpha}\Phi^{[U]}(x,p_{\st}) \nonumber \\
&=& \Big(\Phi_{D}^{\alpha}(x)-\Phi_{A}^{\alpha}(x)\Big)+C_{G}^{[U]}\Phi_{G}^{\alpha}(x) \nonumber \\
&=&\widetilde\Phi_{\partial}^{\alpha}(x) +C_{G}^{[U]}\Phi_{G}^{\alpha}(x).
\label{e:Phip}
\end{eqnarray}
The matrix element $\Phi_{G}^{\alpha}(x)$ is referred to as gluonic pole or Efremov-Teryaev-Qiu-Sterman matrix element~\cite{Efremov:1981sh} and appears multiplied with a gluonic pole prefactor $C_{G}^{[U]}$. All process dependence is isolated in these calculable gluonic pole prefactors. The matrix elements in Eq.~\ref{e:Phip} are defined through~\cite{Buffing:2011mj}
\begin{eqnarray}
&& \Phi_D^\alpha(x) = \int dx_1\ \Phi_D^\alpha(x-x_1,x_1\vert x), \\
&& \Phi_{A}^{\alpha}(x)\equiv \int dx_{1}\,\text{PV}\frac{i}{x_{1}}\,\Phi_{F}^{n\alpha}(x-x_{1},x_{1}|x), \\[0.2cm]
&& \Phi_G^\alpha(x) = \pi\, \Phi^{n\alpha}_{F}(x,0|x), \label{e:PhiG} 
\end{eqnarray}
with
\begin{align}
& \Phi^\alpha_{D\,ij}(x-x_1,x_1\vert x) = \int \frac{d\xi{\cdot}P\,d\eta{\cdot}P}{(2\pi)^2}\,e^{ip_1{\cdot}\eta+i(p-p_1)\cdot \xi}\langle P\vert\overline{\psi}_{j}(0) & \nonumber \\
& \hspace{2.4cm}\times U_{[0,\eta]}\,iD_\st^\alpha(\eta)\,U_{[\eta,\xi]}\psi_{i}(\xi)\vert P\rangle\,\Big|_{LC}, & \label{e:PhiD} \\
& \Phi^{\alpha}_{F\,ij}(x-x_1,x_1\vert x) = \int \frac{d\xi{\cdot}P\,d\eta{\cdot}P}{(2\pi)^2}\,e^{ip_1{\cdot}\eta+i(p-p_1)\cdot \xi}\langle P\vert\overline{\psi}_{j}(0) \nonumber \\
&\hspace{2.4cm}\times U_{[0,\eta]}\,F_\st^{n\alpha}(\eta)\,U_{[\eta,\xi]}\psi_{i}(\xi)\vert P\rangle\,\Big|_{LC}. &
\end{align}
Note a redefinition of the these definitions compared to the Refs.~\cite{Boer:2003cm,Buffing:2012sz,Buffing:2011mj,Buffing:2012rw} regarding factors of $\pi$, which was required for synchronization with the convention used in Ref.~\cite{Buffing:2013kca}. For fragmentation correlators the gluonic poles vanish~\cite{Metz:2002iz}. We therefore do not expect process dependence for the fragmentation correlators and focus on the distribution correlators only.

In the above single transverse weighting example, only one additional operator shows up, i.e. one gluonic pole or partial derivative operator combination. For higher transverse weightings, we get contributions with multiple of such operators in their definition. Anticipating results for transverse weightings with more factors of $p_\st$, we can write down an expansion of the quark correlator as~\cite{Buffing:2012sz}
\begin{align}
& \Phi^{[U]}(x,p_\st) \ =\ \Phi(x,p_\st^2)
+ \frac{p_{\st i}}{M}\,\widetilde\Phi_\partial^{i}(x,p_\st^2)
+ \frac{p_{\st ij}}{M^2}\,\widetilde\Phi_{\partial\partial}^{ij}(x,p_\st^2) & \nonumber \\
& \hspace{4cm}
+ \frac{p_{\st ijk}}{M^3}\,\widetilde\Phi_{\partial\partial\partial}^{\,ijk}(x,p_\st^2) 
+ \ldots & \nonumber \\
& \hspace{1.5cm} + C_{G}^{[U]}\bigg(\frac{p_{\st i}}{M}\,\Phi_{G}^{i}(x,p_\st^2)
+ \frac{p_{\st ij}}{M^2}\,\widetilde\Phi_{\{\partial G\}}^{\,ij}(x,p_\st^2) & \nonumber \\
& \hspace{4cm} + \frac{p_{\st ijk}}{M^3}\,\widetilde\Phi_{\{\partial\partial G\}}^{\,ijk}(x,p_\st^2)
+ \ldots \bigg) & \nonumber \\
& \hspace{1.5cm} + \sum_c C_{GG,c}^{[U]} \bigg(\frac{p_{\st ij}}{M^2}\,\Phi_{GG,c}^{ij}(x,p_\st^2) & \nonumber \\
& \hspace{4cm} + \frac{p_{\st ijk}}{M^3}\,\widetilde\Phi_{\{\partial GG\},c}^{\,ijk}(x,p_\st^2)
+ \dots \bigg) & \nonumber \\
& \hspace{1.5cm} + \sum_c C_{GGG,c}^{[U]}\bigg(\frac{p_{\st ijk}}{M^3}\,\Phi_{GGG,c}^{ijk}(x,p_\st^2) +\ldots \bigg) & \nonumber \\
& \hspace{1.5cm} + \ldots \, , &
\label{e:TMDstructurequarks}
\end{align}
where the index $c$ accounts for the possibility to have multiple ways of tracing the color. Note that there is no summation over $c$ for the single gluonic pole, since only one color structure is allowed in that situation. Contributions like $\widetilde\Phi_{\{\partial G\}}$ indicate the symmetrized combination $\widetilde\Phi_{\{\partial G\}}=\widetilde\Phi_{\partial G}+\widetilde\Phi_{G\partial}$. The important realization is that each of these contributions has a certain behavior under time-reversal symmetry. Contributions with an odd number of gluonic poles are T-odd, whereas all other contributions are T-even. We define the number of operators in the definition of the matrix elements (i.e. the number of gluonic poles and $\partial$'s) as the rank of the matrix element, which equals the number of transverse weightings that is required to obtain the object. Furthermore, as can be seen in Eq.~\ref{e:Phip} for the single weighted case, all process dependence is identified with gluonic pole contributions in the form of prefactors $C_G^{[U]}$, calculable (numerical) factors that depend on the gauge link only.

When performing transverse weightings, we basically weight the expression in Eq.~\ref{e:TMDstructurequarks}. Weighting the correlator $\Phi^{[U]}(x,p_\st)$ with zero factors of $p_\st$ implies that on the r.h.s. of Eq.~\ref{e:TMDstructurequarks} only the object $\Phi(x,p_\st^2)$ survives (or actually the integrated version of it). All other contributions have factors of $p_{\st i}$, $p_{\st ij}$, $p_{\st ijk}$, etc., which do not survive the integration over transverse momentum. Here, these transverse momenta are defined as the symmetric and traceless tensors, e.g.
\begin{equation}
p_{\st i},\, p_{\st ij}=p_{\st i} p_{\st j}-\tfrac{1}{2}p_{\st}^2 g_{\st ij}.
\end{equation}
For a single transverse weighting, we have to multiply Eq.~\ref{e:TMDstructurequarks} with $p_{\st i}$ and integrate over transverse momentum. Due to the definitions of the transverse momentum tensors, on the r.h.s. only the matrix elements with the prefactor $p_{\st i}$ survive, i.e. the integrated versions of $\widetilde\Phi_\partial^{i}(x,p_\st^2)$ and $C_{G}^{[U]}\,\Phi_{G}^{i}(x,p_\st^2)$. This can be generalized for transverse weightings with an arbitrary number of arbitrary rank.

Applying the transverse weightings on TMDs, we obtain the weighted functions
\begin{equation}
f_{\ldots}^{(n)[U]}(x,p_\st^2) = \left(\frac{-p_\st^2}{2M^2}\right)^n\,f_{\ldots}^{[U]}(x,p_\st^2). \label{e:transversemoments}
\end{equation}
Usually only the integrated functions $f_{\ldots}^{(n)[U]}(x)$ are referred to as transverse moment. We will extend this name to functions that still depend on $p_\st^2$, but are azimuthally averaged. The behavior of the TMDs under time reversal symmetry is known. E.g. $f_1$ is T-even, while the Boer-Mulders function $h_1^{\perp[U]}$ is T-odd. We could therefore identify (at the level of transverse moments) which TMD corresponds to which matrix element in the expansion in Eq.~\ref{e:TMDstructurequarks}. For example, $h_1^{\perp}$ corresponds to $C_G^{[U]}\Phi_{G}(x,p_\st^2)$, see e.g. Ref.~\cite{Boer:2003cm}, whereas $f_1$ corresponds to $\Phi(x,p_\st^2)$. This way, all TMDs could be associated with one or more matrix elements. For the rank 2 Pretzelocity function a complication arises, since it corresponds to the matrix elements $\widetilde\Phi_{\partial\partial}^{ij}(x,p_\st^2)$ and $C_{GG,c}^{[U]}\,\Phi_{GG,c}^{ij}(x,p_\st^2)$, the latter coming in two color contributions, see the Refs.~\cite{Buffing:2012sz, Buffing:2012rw}. Therefore, we have three Pretzelocity functions,
\begin{eqnarray}
h_{1T}^{\perp [U]}(x,p_\st^2) & = & h_{1T}^{\perp (A)}(x,p_\st^2) + C_{GG,1}^{[U]}\,h_{1T}^{\perp (B1)}(x,p_\st^2) \nonumber \\
&&+ C_{GG,2}^{[U]}\,h_{1T}^{\perp (B2)}(x,p_\st^2).
\end{eqnarray}
Note that we strictly speaking only make the identification at the level of transverse moments using our methods. Incidentally, for both Drell-Yan and SIDIS we get the same linear combination of them, namely
\begin{equation}
h_{1T}^{\perp [\pm]}(x,p_\st^2) = h_{1T}^{\perp (A)}(x,p_\st^2) + h_{1T}^{\perp (B1)}(x,p_\st^2).
\label{e:pretDYSIDIS}
\end{equation}
Nevertheless, it still is important to realize the underlying structure of these functions.

\section{Gluons}
\label{s:gluons}
\begin{figure}[!tb]
\centering
\includegraphics[width=3.8cm,clip]{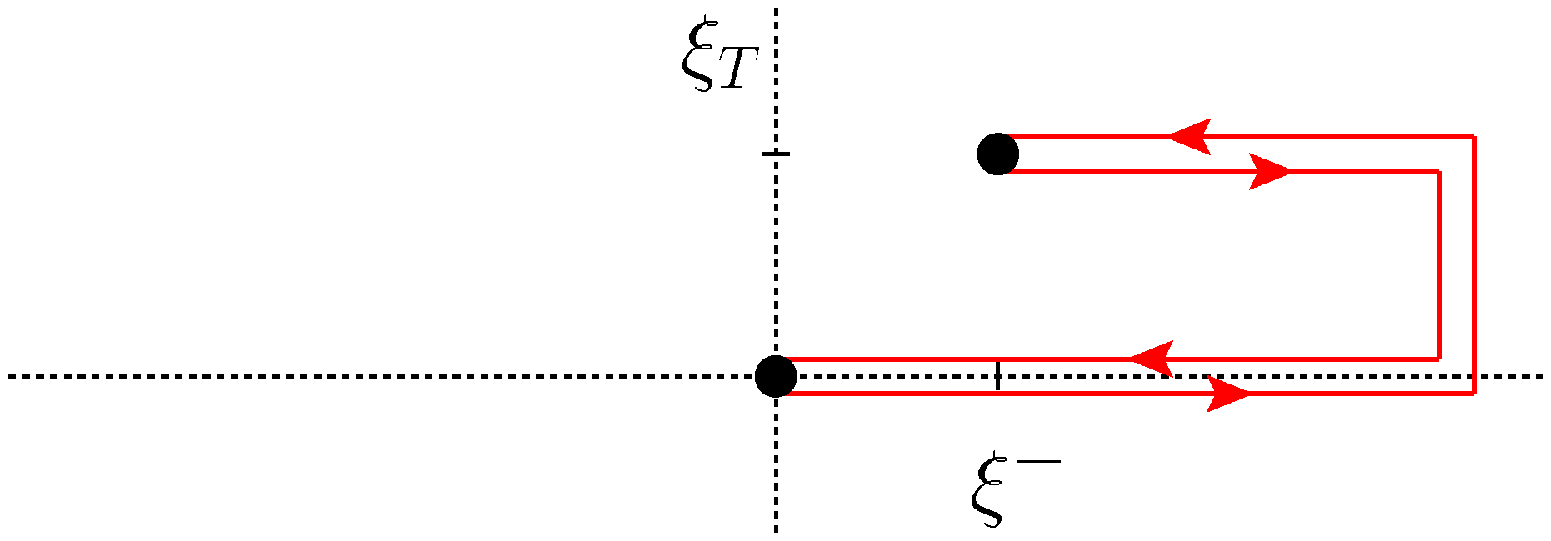}
\hspace{0.3cm}
\includegraphics[width=3.8cm,clip]{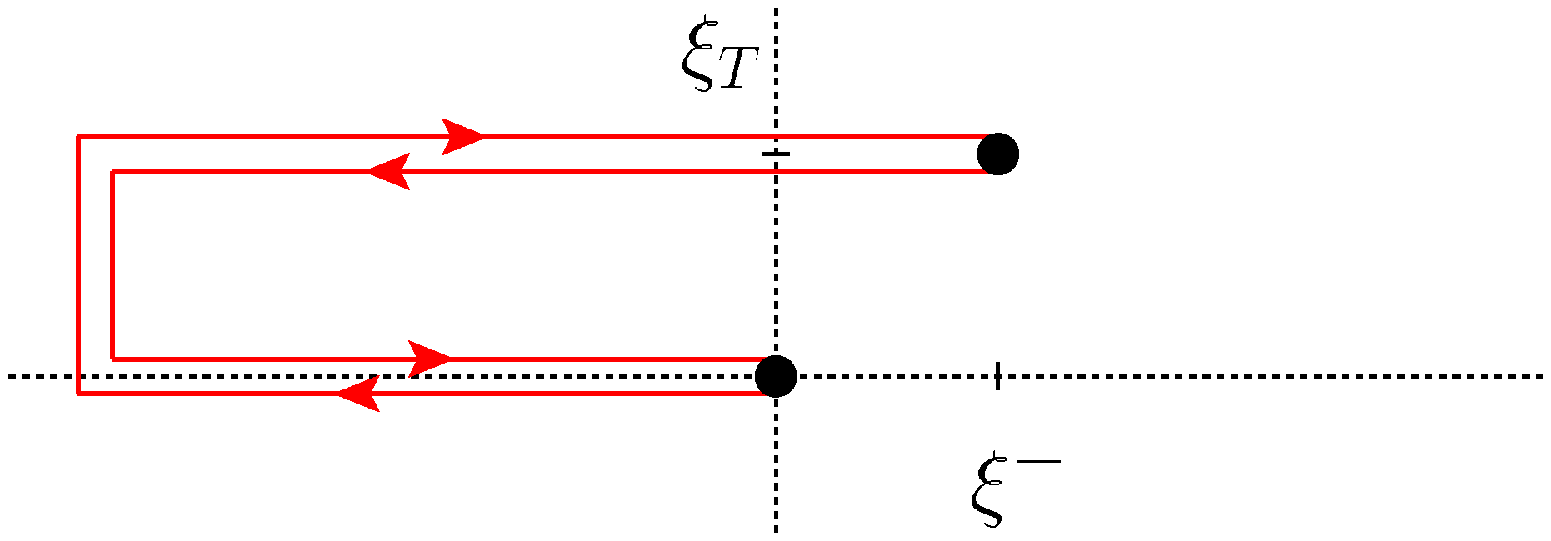}
\\[0.1cm]
\begin{small}
(a)\hspace{3.9cm} (b)
\end{small}
\\[0.3cm]
\includegraphics[width=3.8cm,clip]{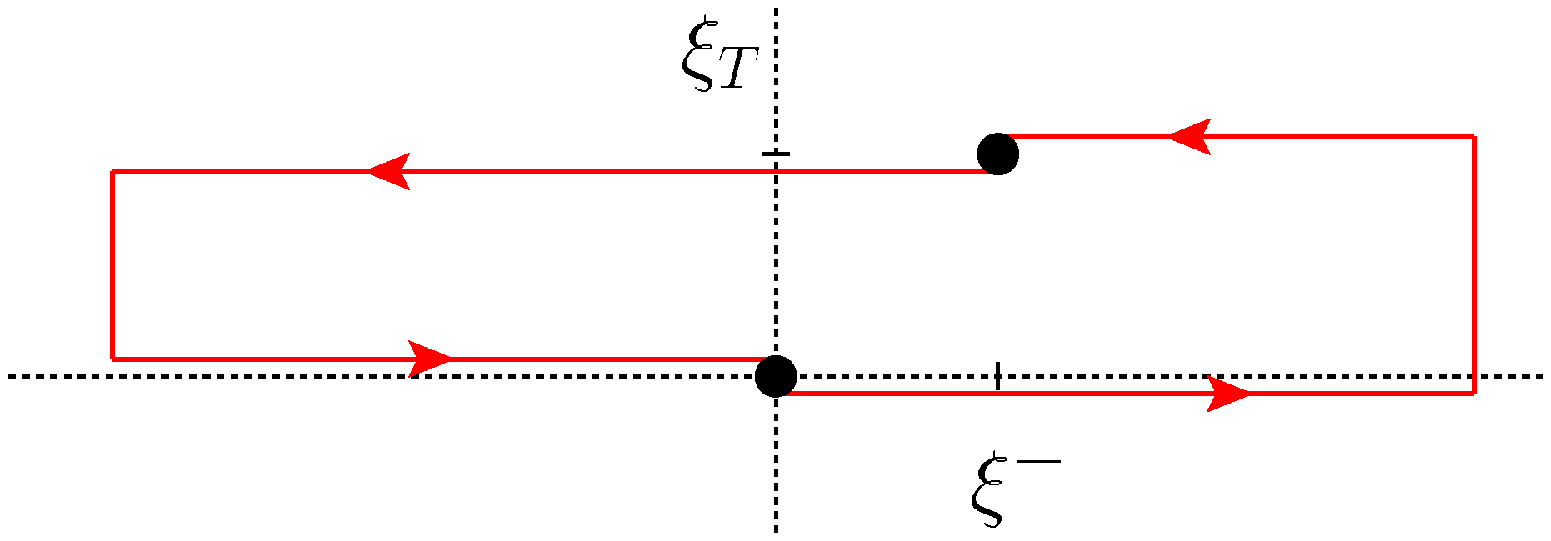}
\hspace{0.3cm}
\includegraphics[width=3.8cm,clip]{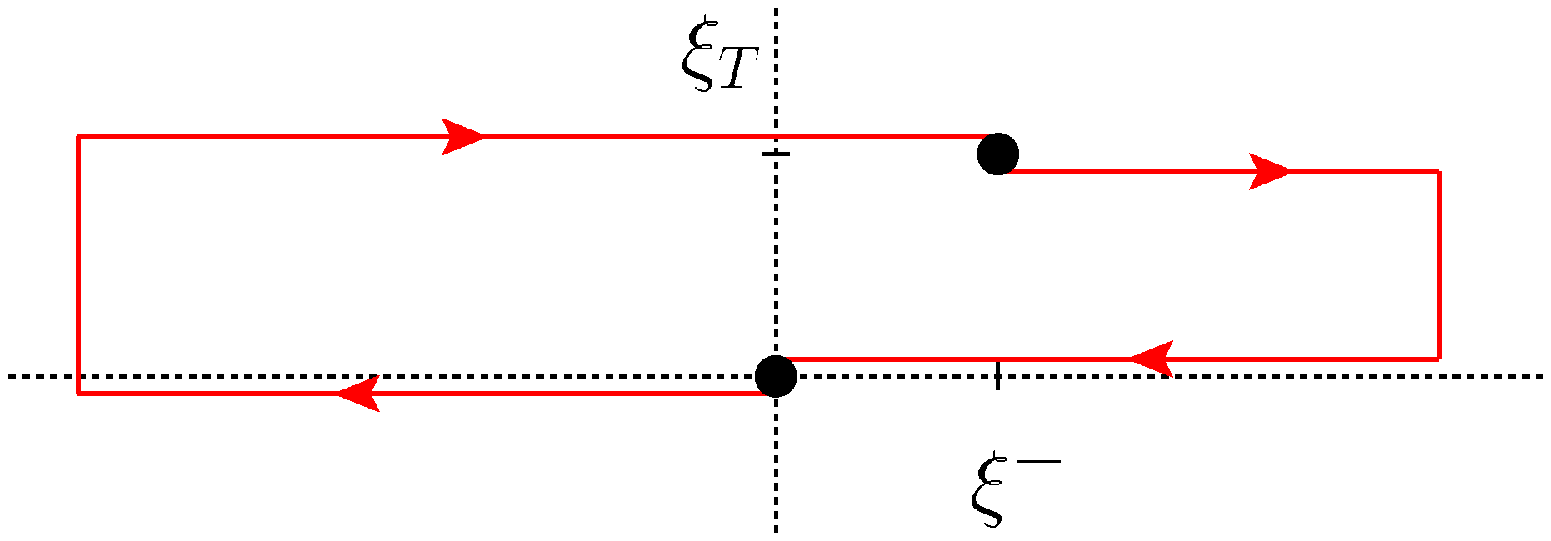}
\\[0.1cm]
\begin{small}
(c)\hspace{3.9cm} (d)
\end{small}
\\[0.3cm]
\includegraphics[width=3.8cm,clip]{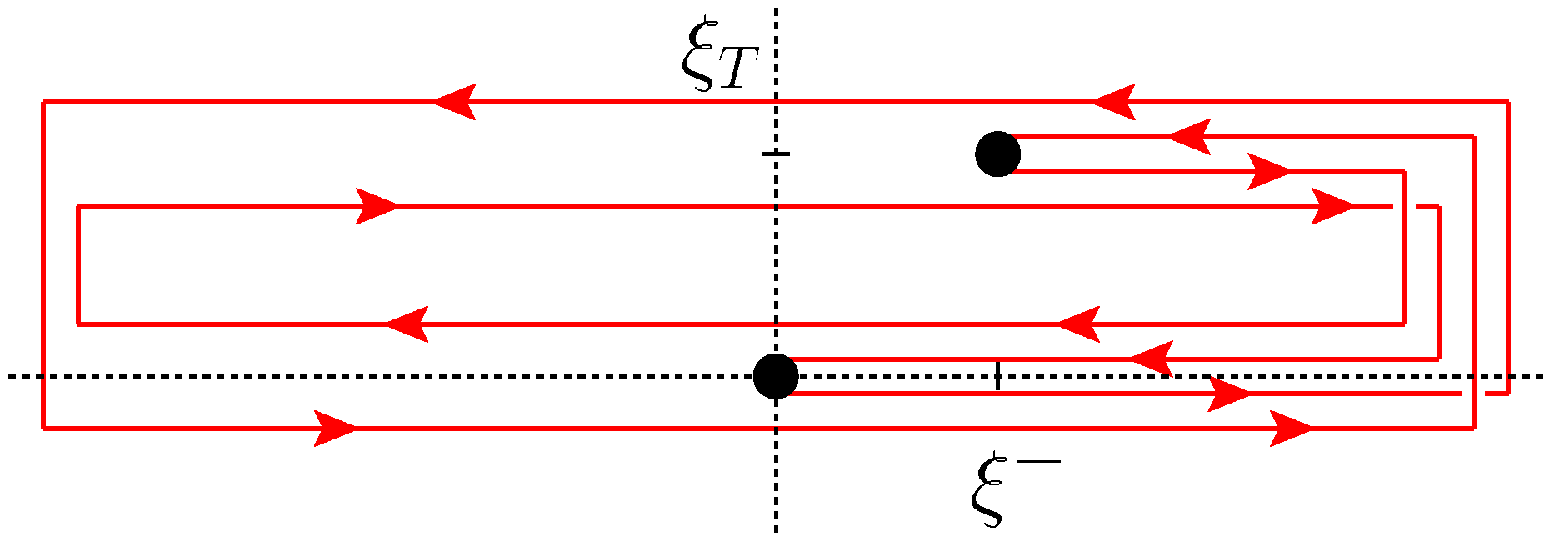}
\hspace{0.3cm}
\includegraphics[width=3.8cm,clip]{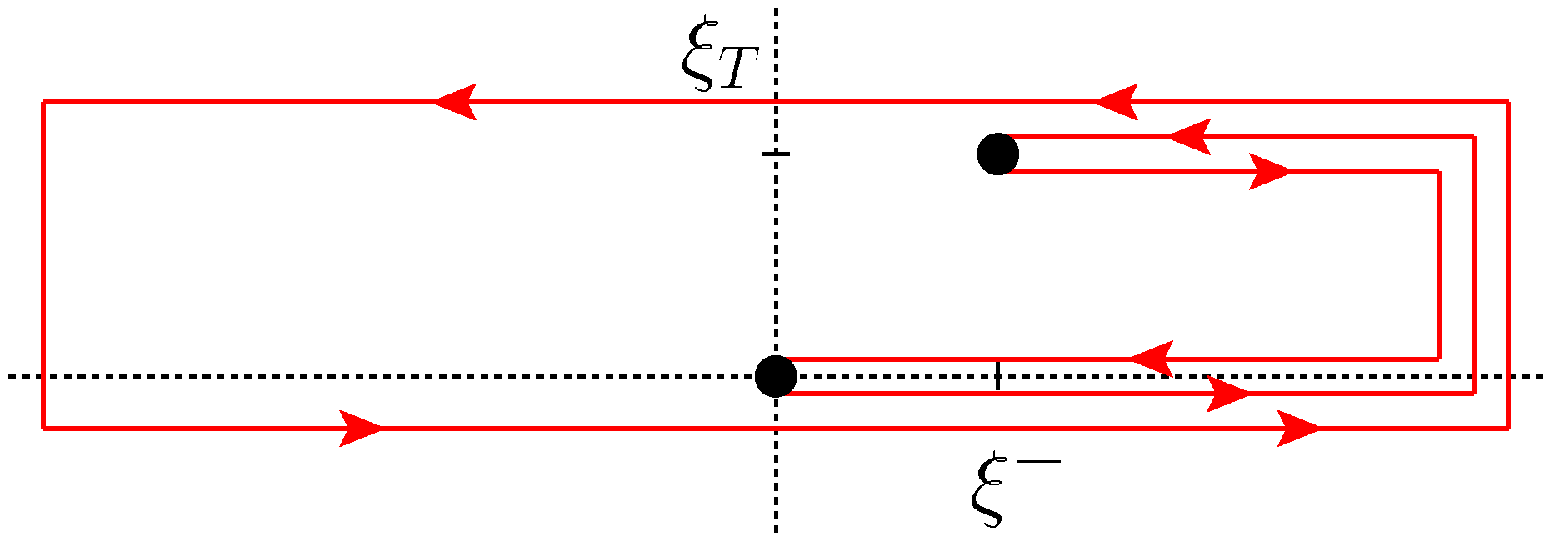}
\\[0.1cm]
\begin{small}
(e)\hspace{3.9cm} (f)
\end{small}
\\[0.3cm]
\includegraphics[width=3.8cm,clip]{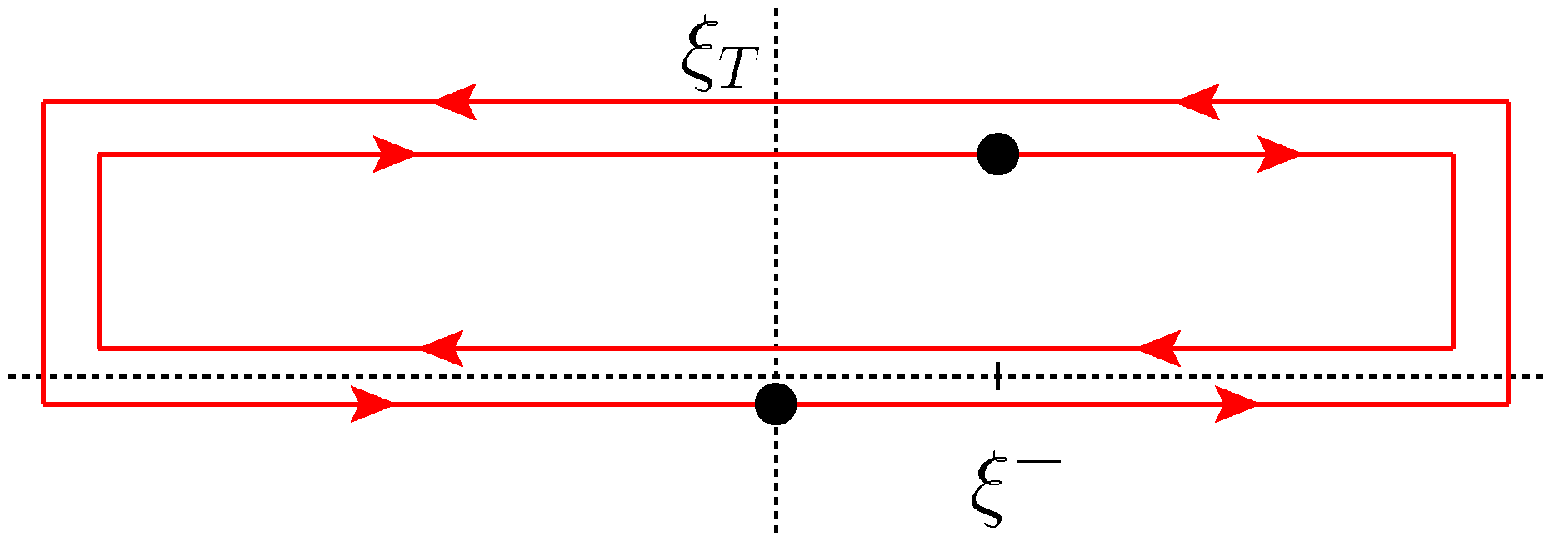}
\hspace{0.3cm}
\includegraphics[width=3.8cm,clip]{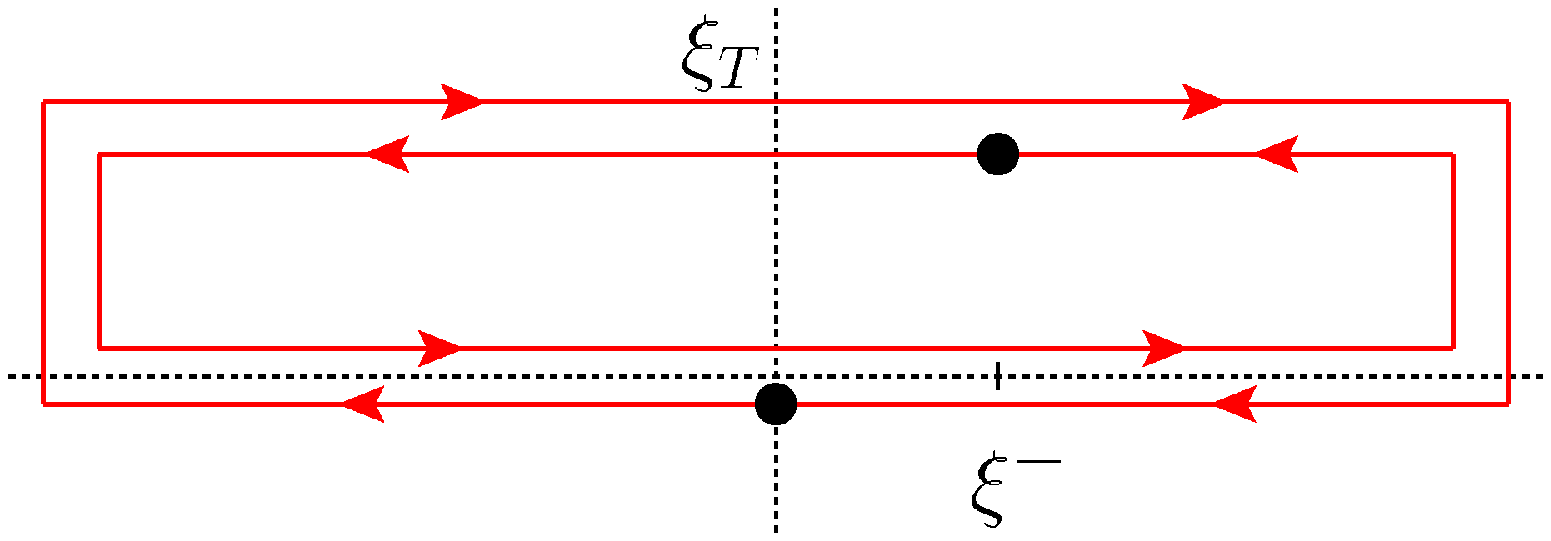}
\\[0.1cm]
\begin{small}
(g)\hspace{3.9cm} (h)
\end{small}
\\[0.1cm]
\caption{Examples of gauge link structures for gluons. The dots represent the locations of the two gluon fields in the gluon correlator, whereas the lines indicate the path of the gauge link. See the main text for an explanation. Figures taken from Ref.~\cite{Buffing:2013kca}.}
\label{f:GL_gluons}
\end{figure}

For gluons, a similar approach can be used and the matrix element for the gluon correlator is given by~\cite{Mulders:2000sh,Bomhof:2006dp,Bomhof:2007xt}
\begin{eqnarray}
\hspace{-7mm}\Gamma^{[U,U^\prime]\,\mu\nu}(x,p_\st;n) & = & {\int}\frac{d\,\xi{\cdot}P\,d^2\xi_\st}{(2\pi)^3}\ e^{ip\cdot\xi}
\,\langle P{,}S|\,F^{n\mu}(0) \nonumber \\
&&\times U_{[0,\xi]}^{\phantom{\prime}}\,F^{n\nu}(\xi)\,U_{[\xi,0]}^\prime\,|P{,}S\rangle\biggr|_{\text{LF}} .
\end{eqnarray}
Note that a color tracing is still required in the above definition. Since the gluon fields are color octets rather than color triplets, two gauge link contributions are required for a proper gauge invariant description, indicated by $U$ and $U^\prime$ in the above equation. Both contributions consist of staple like gauge links, with optionally additional Wilson loops. The three types of structures that can be constructed this way in the relevant $2\rightarrow 2$ processes are given by
\begin{eqnarray}
\hspace{-6mm}\text{type 1:}&&\hspace{-3mm}\tr_c \Big( F^{n\mu}(0)\,U_{[0,\xi]}^{\phantom{\prime}}\,F^{n\nu}(\xi)\,U_{[\xi,0]}^\prime\Big) \nonumber \\
\hspace{-6mm}\text{type 2:}&&\hspace{-3mm}\tr_c \Big( F^{n\mu}(0)\,U_{[0,\xi]}^{\phantom{\prime}}\,F^{n\nu}(\xi)\,U_{[\xi,0]}^\prime\Big) \frac{1}{N_c}\tr_c \Big( U^{[\text{loop}]}\Big) \nonumber \\
\hspace{-6mm}\text{type 3:}&&\hspace{-3mm}\frac{1}{N_c}\tr_c \Big( F^{n\mu}(0)\,U^{[\text{loop}]}\Big)\tr_c \Big( F^{n\nu}(\xi)\,U^{[\text{loop}^{\prime}]}\Big) \nonumber
\end{eqnarray}
The first type corresponds to correlators containing a single color trace only, among them the four simplest gluon gauge link structures allowed, illustrated in Fig.~\ref{f:GL_gluons}(a)-(d). These four gauge link structures consist of the staple links going through plus or minus light cone infinity. Since there are two possibilities for both of them, it leaves us with four structures. More involved structures also allow for e.g. the situation that $U$ and $U^\prime$ are a combination of three staple links, illustrated in more detail in Fig.~\ref{f:GL_gluons}(e). Correlators of the second type have two or more color traces and are extensions of the first type. Starting from the structure of the first type, one can allow for color traces containing gauge link loops only and multiply the type 1 correlator with them, see e.g. Fig.~\ref{f:GL_gluons}(f). In this, we define the gauge link loops as $U^{[\square]}=U_{[0,\xi]}^{[+]}U_{[\xi,0]}^{[-]}$ or $U^{[\square]\dagger}=U_{[0,\xi]}^{[-]}U_{[\xi,0]}^{[+]}$. Type 3 correlators are required too, see Fig.~\ref{f:GL_gluons}(g)-(h), but in these proceedings the focus will be on the type 1 and type 2 color structures.

\begin{figure}[!tb]
\centering
\includegraphics[width=2.5cm,clip]{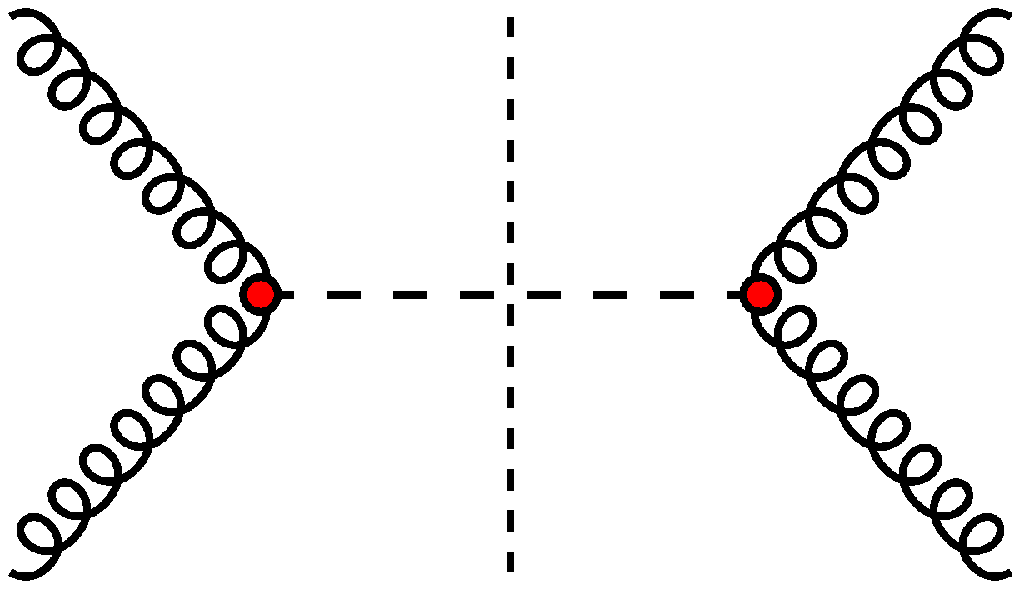}
\hspace{0.15cm}
\includegraphics[width=2.5cm,clip]{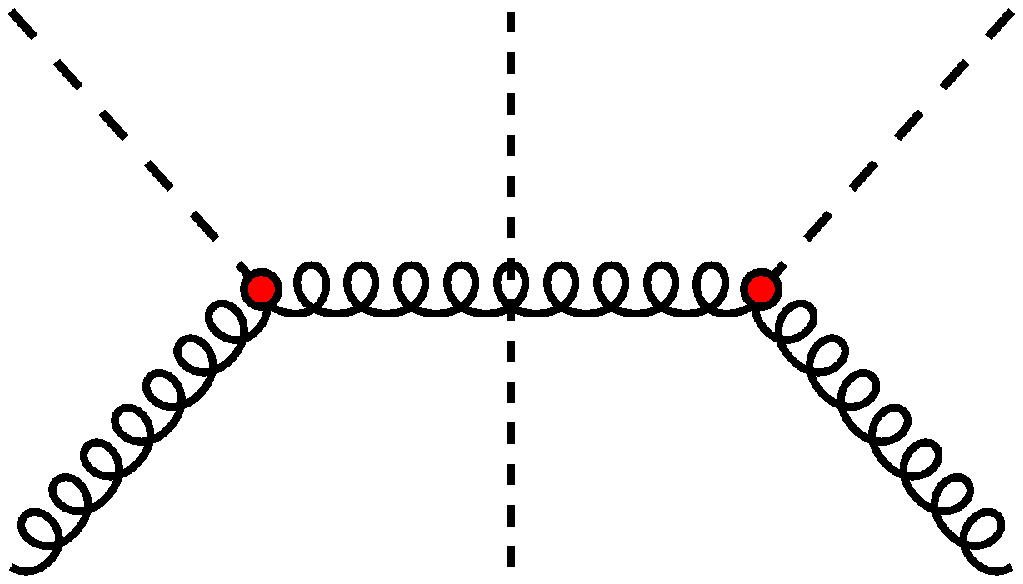}
\hspace{0.15cm}
\includegraphics[width=2.5cm,clip]{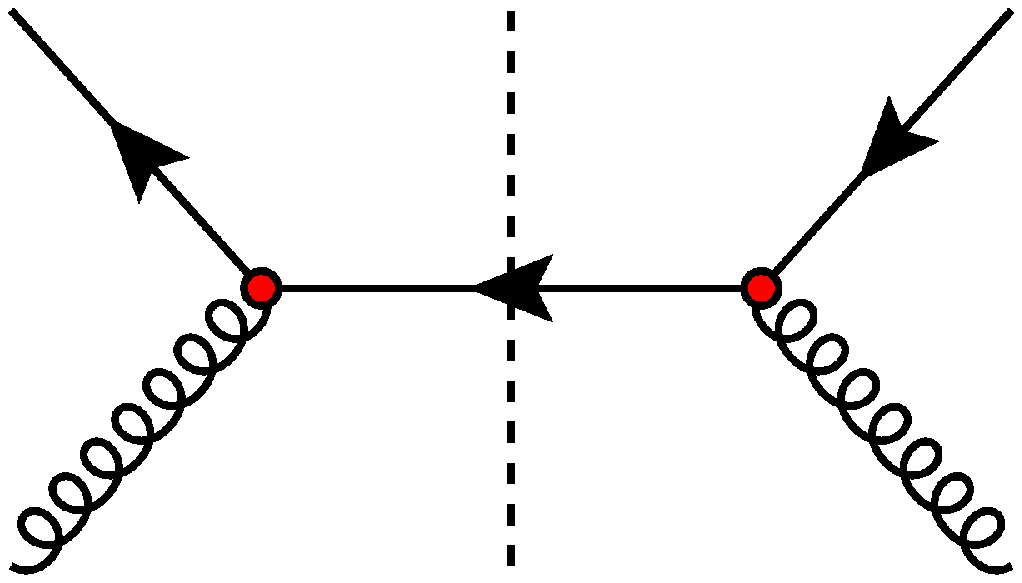}
\\[0.1cm]
\begin{small}
(a) \hspace{2.325cm} (b) \hspace{2.325cm} (c)
\end{small}
\\[0.1cm]
\caption{Three Feynman diagrams with a different color flow contribution each. In (a) all color remains in the initial state, in (b) all color flows into the final state and in (c) we have color splitting, with color flowing in both the initial and final state. See the main text for the implications for the corresponding gauge link structures.}
\label{f:Feyndiag}
\end{figure}

Just as for quarks, the minus gauge links come from initial state interactions and the plus gauge links from final state interactions. A simple illustration for gluons involves the diagrams in Fig.~\ref{f:Feyndiag}. In Fig.~\ref{f:Feyndiag}(a) there are only ISIs and calculations show the gauge link structure to be the one in Fig.~\ref{f:GL_gluons}(b). For Fig.~\ref{f:Feyndiag}(b), with only FSIs, we find the gauge link structure of Fig.~\ref{f:GL_gluons}(a). The Feynman diagram in Fig.~\ref{f:Feyndiag}(c) has color splitting, with color flowing into both the initial and final state. Due to this color structure, we find the gauge link structure in Fig.~\ref{f:GL_gluons}(d), with one staple link running through plus light cone infinity and one staple link running through minus light cone infinity.

On the other hand, TMDs could be used to parametrize the correlator as well, giving for the unpolarized hadron contributions the expression
\begin{eqnarray}
\hspace{-0.7cm} 2x\,\Gamma^{\mu\nu [U]}(x{,}p_\st) &=& -g_T^{\mu\nu}\,f_1^{g [U]}(x{,}p_\st^2) \nonumber \\
&&\hspace{-0.5cm}+\bigg(\frac{p_T^\mu p_T^\nu}{M^2}\,{-}\,g_T^{\mu\nu}\frac{p_\st^2}{2M^2}\bigg)\;h_1^{\perp g [U]}(x{,}p_\st^2),
\label{e:gluonpar}
\end{eqnarray}
where we use the naming convention of Ref.~\cite{Meissner:2007rx}. We refer to Ref.~\cite{Mulders:2000sh} for the full parametrization.

Applying weightings at the level of the matrix elements, one is much more sensitive to the type of gauge link structures involved compared to the quark situation, due to more complicated gauge link structures for gluon correlators. As will be shown later, this results in a much larger set of color combinations of the operator structures. The color index $c$ that for quarks in Eq.~\ref{e:TMDstructurequarks} only became relevant for the Pretzelocity function will play a more significant role for gluons.

Let's start with the contributions for type 1 correlators and focus on the matrix elements containing gluonic poles only. We then have (depending on the gauge link structures involved) the matrix elements
\begin{subequations}
\begin{align}
\Gamma_{G,1}^{\alpha_1}&\rightarrow \tr_c\Big( F(0) \left[G_\st^{\alpha_1}(\xi),F(\xi)\right] \Big)\, , \label{e:colorG1} \\
\Gamma_{G,2}^{\alpha_1}&\rightarrow \tr_c\Big( F(0) \left\{G_\st^{\alpha_1}(\xi),F(\xi)\right\} \Big)\, , \label{e:colorG2} \\
\Gamma_{GG,1}^{\alpha_1\alpha_2}&\rightarrow \tr_c\Big( F(0) \left[G_\st^{\alpha_1}(\xi),\left[G_\st^{\alpha_2}(\xi),F(\xi)\right]\right]\Big)\, , \label{e:colorGG1} \\
\Gamma_{GG,2}^{\alpha_1\alpha_2}&\rightarrow \tr_c\Big( F(0) \left\{G_\st^{\alpha_1}(\xi),\left\{G_\st^{\alpha_2}(\xi),F(\xi)\right\}\right\} \Big)\, , \label{e:colorGG2} \\
\Gamma_{GGG,1}^{\alpha_1\alpha_2\alpha_3}&\rightarrow \tr_c\Big( F(0) \left[G_\st^{\alpha_1}(\xi),\left[G_\st^{\alpha_2}(\xi),\left[G_\st^{\alpha_3}(\xi),F(\xi)\right]\right]\right] \Big)\, , \label{e:colorGGG1} \\
\Gamma_{GGG,2}^{\alpha_1\alpha_2\alpha_3}&\rightarrow \tr_c\Big( F(0) \left\{G_\st^{\alpha_1}(\xi),\left\{G_\st^{\alpha_2}(\xi),\left\{G_\st^{\alpha_3}(\xi),F(\xi)\right\}\right\}\right\} \Big)\, . \label{e:colorGGG2}
\end{align}
\end{subequations}
Note that we omitted the gauge links themselves for the sake of simplicity of this illustration.
Each time, the gluonic poles enter either in a commutator or anticommutator combination with the gluon field $F(\xi)$. For the single weighted case, these functions were introduced in Ref.~\cite{Bomhof:2007xt} with the subscripts $d$ and $f$. For type 2 and type 3 more complicated structures are allowed, since there is an additional color trace that could receive operators due to transverse weighting. Examples of (gluonic pole only) structures that arise are
\begin{subequations}
\begin{align}
\Gamma_{GG,3}^{\alpha_1\alpha_2}&\rightarrow \tfrac{2}{N_c}\tr_c\Big(\left\{ G_\st^{\alpha_1}(\xi), G_\st^{\alpha_2}(\xi)\right\}\Big)\tr_c\Big( F(0)F(\xi)\Big)\, , & \label{e:colorGG3} \\
\Gamma_{GG,4}^{\{\alpha_1\alpha_2\}}&\rightarrow \tfrac{2}{N_c}\tr_c\Big( F(0)\, G_\st^{\{\alpha_1}(\xi)\Big)\tr_c\Big(\big\{ G_\st^{\alpha_2\}}(\xi),F(\xi)\big\}\Big)\, . & \label{e:colorGG4}
\end{align}
\end{subequations}
The parentheses around some indices in some of the above equations indicates a symmetrization over those indices. For contributions receiving $\partial$ contributions only, defined through $i\partial^\alpha=iD_\st^\alpha - A_\st^\alpha$, we find that they always come in the commutator combination, i.e.
\begin{subequations}
\begin{align}
\widetilde\Gamma_{\partial}^{\alpha_1}&\rightarrow \tr_c\Big( F(0) \left[i\partial_{\st}^{\alpha_1},F(\xi)\big]\right] \Big)\, , \label{e:colord} \\
\widetilde\Gamma_{\partial\partial}^{\alpha_1\alpha_2}&\rightarrow \tr_c\Big( F(0) \big[i\partial_{\st}^{\alpha_1}(\xi),\big[i\partial_{\st}^{\alpha_2}(\xi),F(\xi)\big]\big] \Big)\, , \label{e:colordd} \\
\widetilde\Gamma_{\partial\partial\partial}^{\alpha_1\alpha_2\alpha_3}&\rightarrow \tr_c\Big( F(0) \big[i\partial_{\st}^{\alpha_1}(\xi),\big[i\partial_{\st}^{\alpha_2}(\xi),\big[i\partial_{\st}^{\alpha_3}(\xi),F(\xi)\big]\big]\big] \Big)\, . \label{e:colorddd} 
\end{align}
\end{subequations}
On top of this, also a number of mixed terms exists that have both gluonic pole and $\partial$ contributions. A full list of all these contributions (and the gluonic pole contributions not shown above) can be found in Ref.~\cite{Buffing:2013kca}. Again writing down the expansion of the correlator in terms of matrix elements containing gluonic poles and $\partial$ contributions, we find~\cite{Buffing:2013kca,Buffing:2013eka}
\begin{align}
&\Gamma^{[U]}(x,p_\st) \ = \ \Gamma(x,p_\st^2) 
+ \frac{p_{\st i}}{M}\,\widetilde\Gamma_\partial^{i}(x,p_\st^2)
+ \frac{p_{\st ij}}{M^2}\,\widetilde\Gamma_{\partial\partial}^{ij}(x,p_\st^2) & \nonumber \\
& \hspace{4cm}
+ \frac{p_{\st ijk}}{M^3}\,\widetilde\Gamma_{\partial\partial\partial}^{\,ijk}(x,p_\st^2) 
+\ldots & \nonumber \\
& \hspace{1.5cm} 
+ \sum_c C_{G,c}^{[U]}\bigg(\frac{p_{\st i}}{M}\,\Gamma_{G,c}^{i}(x,p_\st^2)
+ \frac{p_{\st ij}}{M^2}\,\widetilde\Gamma_{\{\partial G\},c}^{\,ij}(x,p_\st^2) & \nonumber \\
& \hspace{4cm}
+ \frac{p_{\st ijk}}{M^3}\,\widetilde\Gamma_{\{\partial\partial G\},c}^{\,ijk}(x,p_\st^2) + \ldots\bigg) & \nonumber \\ 
& \hspace{1.5cm}
+ \sum_c C_{GG,c}^{[U]}\bigg(\frac{p_{\st ij}}{M^2}\,\Gamma_{GG,c}^{ij}(x,p_\st^2) & \nonumber \\
& \hspace{4cm} + \frac{p_{\st ijk}}{M^3}\,\widetilde\Gamma_{\{\partial GG\},c}^{\,ijk}(x,p_\st^2) + \ldots\bigg) & \nonumber \\
& \hspace{1.5cm} +
\sum_c C_{GGG,c}^{[U]}\bigg(\frac{p_{\st ijk}}{M^3}\,\Gamma_{GGG,c}^{ijk}(x,p_\st^2)
+ \ldots \bigg) & \nonumber \\
& \hspace{1.5cm} + \ldots \, , &
\label{e:TMDstructuregluons}
\end{align}
The number of color structures, of which some were illustrated above and of which a complete list can be found in Ref.~\cite{Buffing:2013kca}, runs to two for $c=1$, it runs to four for $c=2$ and it runs to seven for $c=3$. Starting from the Eqs.~\ref{e:gluonpar} and \ref{e:TMDstructuregluons}, we can again perform transverse weightings and compare the results of the two separate approaches.

For the approach in terms of matrix elements, we give the double weighted case as an example. We find that
\begin{eqnarray}
\Gamma_{\partial\partial}^{\alpha_1\alpha_2\,[U]}(x) &\equiv & \int d^2p_{\st} \,p_{\st}^{\alpha_1}p_{\st}^{\alpha_2}\,\Gamma^{[U]}(x,p_\st) \nonumber \\
&=&\widetilde\Gamma_{\partial\partial}^{\alpha_1\alpha_2}(x) + \sum_c C_{G,c}^{[U]}\,\widetilde\Gamma_{\{\partial G\},c}^{\alpha_1\alpha_2}(x) \nonumber \\
&& \hspace{10mm} + \sum_c C_{GG,c}^{[U]}\,\Gamma_{GG,c}^{\alpha_1\alpha_2}(x). \label{e:Phiw2}
\end{eqnarray}
We could find this by looking at the r.h.s. of Eq.~\ref{e:TMDstructuregluons}. Only the rank 2 objects on the r.h.s. of that equation survive weighting over two transverse momenta, the reason for which is analogues to the explanation we gave at the end of Section~\ref{s:quarks} for transverse weighting of the quark correlator. Among the surviving matrix elements are specific contributions with zero, one and two gluonic poles which come in different color configurations, hence the summation over the index $c$.

Applying transverse weightings on the TMDs, using the definition in Eq.~\ref{e:transversemoments}, we can identify which TMD corresponds to which matrix element in the expansion in Eq.~\ref{e:gluonpar}. It turns out that $h_{1}^{\perp g}$ is the only gluon TMD contributing at rank 2. It is a T-even function (this function is multiplied by two factors of $p_\st$ in Eq.~\ref{e:gluonpar}) and could therefore correspond to both $\widetilde\Gamma_{\partial\partial}(x,p_\st^2)$ and $C_{GG,c}^{[U]}\,\Gamma_{GG,c}(x,p_\st^2)$, with $c$ running from $1$ to $4$, since there are four possibilities to trace the color. This implies that there are five $h_{1}^{\perp g}$ functions, which depending on the process under consideration appear in different linear combinations, since four of them come with a process dependent gluonic pole factor. There is no identification with $\widetilde\Gamma_{\{\partial G\},c}^{\alpha_1\alpha_2}(x)$, since there are no T-odd rank 2 contributions at leading twist that could be identified with it. Including the results for the TMDs not explicitly mentioned in Eq.~\ref{e:gluonpar}, this leads for the gluon TMDs to the results
\begin{eqnarray}
f_{1T}^{\perp g[U]}(x,p_\st^2)&=&\sum_{c=1}^2 C_{G,c}^{[U]}\,f_{1T}^{\perp g(Ac)}(x,p_\st^2), \label{e:f1Tperpg} \\
h_{1T}^{g[U]}(x,p_\st^2)&=&\sum_{c=1}^2 C_{G,c}^{[U]}\,h_{1T}^{g(Ac)}(x,p_\st^2), \label{e:h1Tg} \\
h_{1L}^{\perp g[U]}(x,p_\st^2)&=&\sum_{c=1}^2 C_{G,c}^{[U]}\,h_{1L}^{\perp g(Ac)}(x,p_\st^2), \label{e:h1Lperpg} \\
h_1^{\perp g[U]}(x,p_\st^2)&=&h_1^{\perp g (A)}(x,p_\st^2) \nonumber \\
&&+\sum_{c=1}^{4}C_{GG,c}^{[U]}\,h_1^{\perp g (Bc)}(x,p_\st^2), \label{e:h1perpg} \\
h_{1T}^{\perp g[U]}(x,p_\st^2)&=&\sum_{c=1}^2 C_{G,c}^{[U]}\,h_{1T}^{\perp g(Ac)}(x,p_\st^2) \nonumber \\
&&+\sum_{c=1}^{7}C_{GGG,c}^{[U]}\,h_{1T}^{\perp g(Bc)}(x,p_\st^2). \label{e:h1Tperpg}
\end{eqnarray}
The three TMDs not mentioned above in the Eqs.~\ref{e:f1Tperpg}~-~\ref{e:h1Tperpg}, namely $f_1^g$, $g_1^g$ and $g_{1T}^g$, are process independent.

To illustrate this generalized universality for the $h_1^\perp$, consider the situations for the three diagrams illustrated in Fig.~\ref{f:Feyndiag}. We find for both the Higgs production through gluon fusion and the scattering of a gluon on a Higgs particle that
\begin{subequations}
\begin{align}
h_1^{\perp g[-,-]}(x,p_\st^2)&=h_1^{\perp g (A)}(x,p_\st^2)+h_1^{\perp g (B1)}(x,p_\st^2), & \\
h_1^{\perp g[+,+]}(x,p_\st^2)&=h_1^{\perp g (A)}(x,p_\st^2)+h_1^{\perp g (B1)}(x,p_\st^2), &
\end{align}
\end{subequations}
whereas we find for the color splitting example in Fig.~\ref{f:Feyndiag}(c) that
\begin{equation}
h_1^{\perp g[-,+]}(x,p_\st^2)=h_1^{\perp g (A)}(x,p_\st^2)+h_1^{\perp g (B2)}(x,p_\st^2).
\end{equation}
In order to find the functions $h_1^{\perp g (B3)}(x,p_\st^2)$ and $h_1^{\perp g (B4)}(x,p_\st^2)$ more complicated diagrams have to be considered.

\section{Conclusions}
\label{s:Conclusions}
For the quarks, a result of applying the method of generalized universality is the discovery of three Pretzelocity functions rather than one. In any process in particular it is a linear combination of these functions that appears. It is the gauge link structure of the diagram under consideration that determines which linear combination appears. For Drell-Yan and SIDIS one does find the same linear combination of Pretzelocity functions. Nevertheless it is still important to know the precise operator structure underlying the TMDs, since it is important for studies wherein the operator structures involved become relevant, e.g. in lattice calculations.

For the gluon TMDs $f_{1T}^{\perp g[U]}$, $h_{1T}^{g[U]}$, $h_{1L}^{\perp g[U]}$, $h_1^{\perp g[U]}$ and $h_{1T}^{\perp g[U]}$ multiple functions appear and linear combinations of these functions have to be considered. This brings the number of TMDs operator-wise at 23, although there still are only 8 observable TMD structures. Nevertheless, it can be calculated how each of these observable structures are constructed out of the 23 objects for any given process.

\begin{acknowledgement}
This conference proceeding contribution is based on the talk given by MGAB. This research is part of the research program of the ``Stichting voor Fundamenteel Onderzoek der Materie (FOM)'', which is financially supported by the ``Nederlandse Organisatie voor Wetenschappelijk Onderzoek (NWO)''. We also acknowledge support of the FP7 EU-programme HadronPhysics3 (contract no 283286) and QWORK (contract 320389).
\end{acknowledgement}


\begin{thebibliography}{88}

\bibitem{Ralston:1979ys}
J.P.\ Ralston and D.E.\ Soper, Nucl.\ Phys.\ \textbf{B~152}, 109 (1979); %.
%\bibitem{Tangerman:1994eh}
R.D.\ Tangerman and P.J.\ Mulders, Phys.\ Rev.\ \textbf{D~51}, 3357-3372 (1995); %.
%\bibitem{Boer:1999mm}
D.\ Boer, Phys.\ Rev.\ \textbf{D~60}, 014012 (1999).

\bibitem{Buffing:2012sz}
M.G.A.\ Buffing, A.\ Mukherjee and P.J.\ Mulders, Phys.\ Rev.\ \textbf{D~86}, 074030 (2012).

\bibitem{Buffing:2013kca}
M.G.A.\ Buffing, A.\ Mukherjee and P.J.\ Mulders, Phys.\ Rev.\ \textbf{D~88}, 054027 (2013).

\bibitem{Bomhof:2006dp}
C.J.\ Bomhof, P.J.\ Mulders and F.\ Pijlman, Eur.\ Phys.\ J.\ \textbf{C~47}, 147-162 (2006).

\bibitem{Sivers:1989cc}
D.W.\ Sivers, Phys.\ Rev.\ \textbf{D~41}, 83 (1990); %.
%\bibitem{Sivers:1990fh}
D.W.\ Sivers, Phys.\ Rev.\ \textbf{D~43}, 261-263 (1991); %.
%\bibitem{Collins:1992kk}
J.C.\ Collins, Nucl.\ Phys.\ \textbf{B~396}, 161-182 (1993); %.
%\bibitem{Collins:2002kn}
J.C.\ Collins, Phys.\ Lett.\ \textbf{B~536}, 43-48 (2002); %.
%\bibitem{Brodsky:2002rv}
S.J.\ Brodsky, D.S.\ Hwang and I.\ Schmidt, Nucl.\ Phys.\ \textbf{B~642}, 344-356 (2002).

\bibitem{Bacchetta:2006tn}
A.\ Bacchetta, M.\ Diehl, K.\ Goeke, A.\ Metz, P.J.\ Mulders and M.\ Schlegel, JHEP\ \textbf{0702}, 093 (2007).

\bibitem{Boer:2003cm}
D.\ Boer, P.J.\ Mulders and F.\ Pijlman, Nucl.\ Phys.\ \textbf{B~667} 201-241 (2003); %.
%\bibitem{Bacchetta:2005rm}
A.\ Bacchetta, C.J.\ Bomhof, P.J.\ Mulders and F.\ Pijlman, Phys.\ Rev.\ \textbf{D~72}, 034030 (2005); %.
%\bibitem{Bomhof:2006ra}
C.J.\ Bomhof and P.J.\ Mulders, JHEP\ \textbf{0702}, 029 (2007).

\bibitem{Efremov:1981sh}
A.V.\ Efremov and O.V.\ Teryaev, Sov.\ J.\ Nucl.\ Phys.\ \textbf{36}, 140 (1982); %.
%\bibitem{Efremov:1984ip}
A.V.\ Efremov and O.V.\ Teryaev, Phys.\ Lett.\ \textbf{B~150}, 383 (1985); %.
%\bibitem{Qiu:1991pp}
J-W.\ Qiu and G.F.\ Sterman, Phys.\ Rev.\ Lett.\ \textbf{67}, 2264-2267 (1991); %.
%\bibitem{Qiu:1991wg}
J-W.\ Qiu and G.F.\ Sterman, Nucl.\ Phys.\ \textbf{B~378}, 52-78 (1992); %.
%\bibitem{Qiu:1998ia}
J-W.\ Qiu and G.F.\ Sterman, Phys.\ Rev.\ \textbf{D~59}, 014004 (1998); %.
%\bibitem{Kanazawa:2000hz}
Y.\ Kanazawa and Y.\ Koike, Phys.\ Lett.\ \textbf{B~478}, 121-126 (2000).

\bibitem{Buffing:2011mj}
M.G.A.\ Buffing and P.J.\ Mulders, JHEP\ \textbf{1107}, 065 (2011).

\bibitem{Buffing:2012rw}
M.G.A.~Buffing and P.J.~Mulders, Int.\ J.\ Mod.\ Phys.\ Conf.\ Ser.\ \textbf{20}, 66 (2012).

\bibitem{Metz:2002iz}
A.\ Metz, Phys.\ Lett.\ \textbf{B~549}, 139-145 (2002); %.
%\bibitem{Collins:2004nx}
J.C.\ Collins and A.\ Metz, Phys.\ Rev.\ Lett.\ \textbf{93}, 252001 (2004); %.
%\bibitem{Gamberg:2008yt}
L.P.\ Gamberg, A.\ Mukherjee and P.J.\ Mulders, Phys.\ Rev.\ \textbf{D~77}, 114026 (2008); %.
%\bibitem{Meissner:2008yf}
S.\ Meissner and A.\ Metz, Phys.\ Rev.\ Lett.\ \textbf{102}, 172003 (2009); %.
%\bibitem{Gamberg:2010uw}
L.P.\ Gamberg, A.\ Mukherjee and P.J.\ Mulders, Phys.\ Rev.\ \textbf{D~83}, 071503 (2011).

\bibitem{Mulders:2000sh}
P.J.\ Mulders and J.\ Rodrigues, Phys.\ Rev.\ \textbf{D~63}, 094021 (2001).

\bibitem{Bomhof:2007xt}
C.J.\ Bomhof and P.J.\ Mulders, Nucl.\ Phys.\ \textbf{B~795}, 409-427 (2008).

\bibitem{Meissner:2007rx}
S.\ Meissner, A.\ Metz and K.\ Goeke, Phys.\ Rev.\ \textbf{D~76}, 034002 (2007).

\bibitem{Buffing:2013eka}
M.G.A.\ Buffing, P.J.\ Mulders and A.\ Mukherjee, Int.\ J.\ Mod.\ Phys.\ Conf.\ Ser.\ \textbf{25}, 1460003 (2014).

\end{thebibliography}
\end{document}